 \documentclass{jpp}
 \usepackage{epstopdf, epsfig}
 \usepackage{hyperref}
 \usepackage[utf8]{inputenc}

 \shorttitle{Combined plasma-coil optimization algorithms}
 \shortauthor{S. A. Henneberg, S. R. Hudson, D. Pfefferlé, P. Helander}

 \title{Combined plasma-coil optimization algorithms}

 \author{S. A. Henneberg\aff{1}
  \corresp{\email{Sophia.Henneberg@ipp.mpg.de}},
  S. R. Hudson\aff{2},
  D. Pfefferlé\aff{3},
 \and P. Helander\aff{1}}

\affiliation{\aff{1}Max-Planck-Institut f\"{u}r Plasmaphysik, Wendelsteinstr. 1, 17489 Greifswald
\aff{2}Princeton Plasma Physics Laboratory, P.O. Box 451, Princeton, NJ 08543, USA
\aff{3}The University of Western Australia, 35 Stirling Highway, Crawley, WA 6009, Australia}

 \usepackage{graphicx}
 \usepackage{eufrak}
 \usepackage{bm}
 \usepackage{amsfonts}
 \usepackage{amsmath}
 \usepackage{mathtools}
 \usepackage{xcolor}
 \usepackage{pifont} 
 \usepackage{cancel} 
 \usepackage{soul} 

 \definecolor{red}{rgb}{1,0,0}
 \definecolor{gre}{rgb}{0,1,0}
 \definecolor{blu}{rgb}{0,0,1}

\newif\ifcomment 
\commentfalse

\begin{document}
	
 \newcommand{\iotabar}{\text{$\iota\!\!$-}}
	
	\newcommand*\diff{\mathop{}\!\mathrm{d}} 
	\newcommand*\Diff[1]{\mathop{}\!\mathrm{d^#1}}
	\newcommand{\curl}{\nabla\times}
	\renewcommand{\a}{\alpha}
	
	\newcommand{\bb}{\bm{B}}
	\newcommand{\be}{\bm{e}}
	\newcommand{\bMt}{{\sf M}}
	\newcommand{\bxi}{\bm{\xi}}
	\newcommand{\br}{\bm r}
	\newcommand{\bR}{{\sf R}}
	\newcommand{\bphi}{\bm \phi}
	\newcommand{\bkappa}{\bm \kappa}
	\newcommand{\bn}{\bm n}
	\newcommand{\bx}{\bm x}
	\newcommand{\bk}{\bm K}
	\newcommand{\bs}{\bm{s}}
	\newcommand{\bj}{\bm j}
	\newcommand{\bi}{{\sf I}} 
	\newcommand{\ba}{\bm A}
	\newcommand{\vac}{\text{vac}}

 \newcommand{\cA}{{\cal A}}
 \newcommand{\cD}{{\cal D}}
 \newcommand{\MB}{{\cal D}}
 \newcommand{\cF}{{\cal F}}
 \newcommand{\cV}{{\cal V}}
 \newcommand{\pcV}{{\cal S}}
 \newcommand{\cI}{{\cal I}}
 \newcommand{\BTn}{B_{T,n}}
 \newcommand{\BPn}{B_{P,n}}
 \newcommand{\BEn}{B_{E,n}}

 \newcommand{\bRx}{ {\bm{r}}_{\xi}}
 \newcommand{\rrr}{(\bm{r}/r^3)}
 \newcommand{\rrrL}{(\br/r^3)}
 \newcommand{\bart}{{\bar\theta}}
 \newcommand{\barz}{{\bar\zeta}}

 \newcommand{\bmB}{\bm{B}}
 \newcommand{\bfs}{\bm{s}}
 \newcommand{\bfr}{\bm{r}}
 \newcommand{\bfe}{\bm{e}}
 \newcommand{\bmx}{\bm{x}}
 \newcommand{\bmc}{\bm{c}}
 \newcommand{\bmz}{\bm{z}}
	
 \newcommand{\boldmu}{\mbox{\boldmath$\mu$\unboldmath}}
 \newcommand{\boldpsi}{\mbox{\boldmath$\psi$\unboldmath}}
 \newcommand{\boldPsi}{\mbox{\boldmath$\Psi$\unboldmath}}
 \newcommand{\calF}{{\cal F}}
 \newcommand{\calV}{{\cal V}}
 \newcommand{\calE}{{\cal E}}
 \newcommand{\calH}{{\cal H}}
 \newcommand{\calD}{{\cal D}}
 \newcommand{\ds}{\displaystyle}
 \newcommand{\beitem}{\begin{enumerate}}
 \newcommand{\eeitem}{\end{enumerate}}
 
 \newcommand{\Eqn}[1]{Eqn.~\ref{eqn:#1}}
 \newcommand{\Sec}[1]{Sec.~\ref{sec:#1}}
 \newcommand{\App}[1]{App.~\ref{app:#1}}

	\newcommand{\cP}{\mathcal{P}} 
	\newcommand{\cVe}{\mathcal{V}_2} 
	\newcommand{\cVtot}{\mathcal V} 
	
 \maketitle
	
 \begin{abstract}
 Combined plasma-coil optimization approaches for designing stellarators are discussed and a new method for calculating free-boundary equilibria is proposed.
 Four distinct categories of stellarator optimization, two of which are novel approaches, are the fixed-boundary optimization, the generalized fixed-boundary optimization, the quasi free-boundary optimization, and the free-boundary (coil) optimization.
 These are described using the multi-region relaxed magnetohydrodynmics (MRxMHD) energy functional, the Biot-Savart integral, the coil-penalty functional and the virtual casing integral, and their derivatives.
 The proposed free-boundary equilibrium calculation differs from existing methods in how the boundary-value problem is posed, and for the new approach it seems that there is not an associated energy minimization principle because a non-symmetric functional arises.
 We propose to solve the weak formulation of this problem using a spectral-Galerkin method, and this will reduce the free-boundary equilibrium calculation to something comparable to a fixed-boundary calculation. 
 In our discussion of combined plasma-coil optimization algorithms, we emphasize the importance of the stability matrix.
 \end{abstract}

 \section{Introduction}
 
 The design space for stellarators is larger than that of tokamaks because stellarators exploit {\it three-dimensional} (3D) magnetic fields, by which it is meant that there is no continuous (e.g. rotational) symmetry, whereas tokamaks are notionally {\it axisymmetric} (2D) \citep{Helander14}.
 The larger design space allows more freedom in the geometry of the plasma boundary.
 Geometry affects several important plasma properties such as stability and transport \citep{Helander2012}, including turbulent transport \citep{Hegna2018}.
 If the search includes 3D configurations, generally, one may expect that it is more likely to find a feasible fusion device.
  
 A particularly advantageous feature of stellarators is that the rotational-transform, which is essential for confinement in toroidal configurations, can be provided externally \citep{Mercier1964}, either by current-carrying coils or by permanent magnets.
 This reduces or even eliminates the need of generating toroidal plasma currents, which can lead to problematic disruptions.
 
 Along with advantages, 3D configurations can have drawbacks.
 A feasible fusion device must possess good particle confinement and the plasma equilibrium must be supported by an external set of coils or magnets that do not pose unrealistic engineering challenges  \citep{Grieger1992}.
 Stellarators are guaranteed neither.
 Plasmas in early stellarator designs were not well confined because of neoclassical particle losses caused by unconfined orbits \citep{Galeev1969,Beidler2011}.  
 Geometrically distorted coils complicated the engineering of the cancelled NCSX experiment \citep{NCSXrisk} and caused delays in the construction of Wendelstein 7-X~\citep{W7Xexp}. 
 
 Nonetheless, with careful exploitation of the large design space of 3D configurations confinement in stellarators can be significantly improved, e.g., by using quasi-symmetric \citep{Nuhrenberg1988,Boozer1995,Henneberg2020} or quasi-isodynamic fields \citep{Quasi-isodyanmic,Omnigenity}. 
 Perfectly quasi-isodynamic fields have vanishing bootstrap current, which allows for simultaneous optimization of both neoclassical transport and small toroidal net current \citep{Helander2009}, which can be beneficial for stability and is necessary if an island divertor is to be employed.
 The 3D optimization effort is encouraged by recent Wendelstein 7-X results \citep{Klinger2019}, which was optimized for neoclassical transport \citep{Nuehrenberg1986}. 
 There are new methods in coil-design, such as using permanent magnets \citep{PermanentMagnets,Zhu2020,Landreman2020} and designing coils with generous tolerances \citep{Lobsien2018,Lobsien2020}.
 
 With these developments in understanding and designing 3D configurations, the question is not whether we {\it should} search the large 3D configuration space but {\it how}.
 With an order of magnitude more degrees of freedom, give or take, it is significantly more difficult to search the 3D configuration space.
 To simultaneously achieve the desired properties of a feasible stellarator, we need efficient optimization approaches, and we need to understand the solution space.

 An essential component of stellarator optimization is the evaluation of equilibrium properties.\footnote{We consider vacuum approximation or large aspect ratio expansions to be types of equilibrium calculations.} 
 Equilibrium calculations are conveniently divided into two types, namely fixed-boundary and free-boundary.
 \footnote{Here the word ``boundary'' refers to the {\it plasma} boundary; later we shall generalize this terminology to include equilibrium calculations with a fixed {\it computational} boundary but with a plasma boundary that moves during the calculation.}
 The fixed-boundary approach requires the geometry of the plasma boundary to be provided as input information \citep{VMEC,Bauer1984,Hudson2012}.
 Experience suggests that fixed-boundary calculations are faster and more robust than their free-boundary counterpart.
 In reality, however, the geometry of the plasma boundary is not known {\it a priori}.
 Free-boundary equilibrium calculations require as input the external magnetic field, and the self-consistent plasma boundary is determined as part of the equilibrium calculation \citep{fbVMEC, Hudson2020}.
 Existing free-boundary equilibrium codes invariably perform additional iterations compared to their fixed-boundary analogs. 
 The present algorithm in free-boundary SPEC \citep{Hudson2020}, for example, performs a Picard-style iteration over the fixed-boundary code in order to determine the plasma boundary that is consistent with the external field.
 The speed of existing fixed-boundary codes is advantageous for stellarator optimization because the equilibrium is typically computed at each iteration.
 
 One might ask, what is the ``best'' approach for stellarator optimization?
 We do not expect that there will be a be-all and end-all of stellarator optimization algorithms.
 We should embrace a variety of methods.
 In this paper, we seek to elucidate the mathematical structure of the various inter-related calculations that underpin the integrated stellarator optimization problem, which we hereafter call the combined plasma-coil optimization.
 The overall numerical efficiency of the combined plasma-coil optimization problem cannot be understood without considering how the equilibrium calculation, the coil calculation, and the optimization calculation communicate with each other.
 This needs to be understood from both a mathematical and an algorithmic perspective.
 
 Existing algorithms can loosely be sorted into two categories, which may be called the ``two-step'' method and the ``direct-coil'' method, which we now describe.%
 \footnote{Note that this is loose terminology and there are numerous variations of these methods. We shall give a more precise description of various algorithms in \Sec{optimization}.}
 The two-step approach separates the equilibrium optimization calculation from the problem of finding coils \citep{Nuehrenberg1986}.
 In the first step, the plasma boundary is varied to optimize the desired equilibrium properties, such as rotational transform profile, MHD stability, etc. \citep{Beidler1990,Zarnstorff2001,Strickler2002, Henneberg19, Bader2020}.
 In the second step, the geometry of an external set of coils that provides the required external field is determined. 
 The coil geometry must meet certain engineering criteria; the coils cannot have too small curvature and they must not intersect each other, for example.
 This stellarator optimization can be performed with STELLOPT \citep{STELLOPT}, ROSE \citep{Drevlak2019}, or SIMSOPT \citep{SIMSOPT}, which are all under active development.
 The two-step approach was used to design Wendelstein 7-X \citep{Beidler1990}, CHS-qa \citep{Okamura2001}, NCSX~\citep{Nelson2003}, HSX \citep{HSX07}, ESTELL \citep{ESTELL}, and CFQS \citep{Shimizu2018}.

 An advantage of the two-step approach is that it is primarily the fusion-relevant performance of the plasma that is most important; so, it is reasonable to optimize the plasma performance first without compromise; that is, without regard to the complexity of the coils.
 After all, a stellarator that produces fusion power with complicated coils is incomparably more commercially valuable than a stellarator with simple coils that does not.
 An advantage of using fixed-boundary calculations to optimize the equilibrium properties is the availability of fast and robust fixed-boundary equilibrium codes. 
 
 Using fixed-plasma-boundary calculations for the plasma equilibrium leads to an inverse problem for the coil geometry: the required magnetic field is known, and the Biot-Savart law must be {\it inverted} to obtain the coil geometry.
 Many different codes have been developed to solve the inverse problem for the coils, e.g.,  NESCOIL~\citep{NESCOIL}, ONSET~\citep{ONSET},  COILOPT~\citep{COILOPT}, FOCUS~\citep{FOCUS1},  REGCOIL~\citep{REGCOIL}, OMIC~\citep{Singh2020}, and FOCUSADD~\citep{McGreivy2020}.
 There are codes that solve the inverse problem for a set of permanent magnets \citep{PermanentMagnets,Zhu2020,Landreman2020}.

 A disadvantage of the two-step approach is that not all plasma boundaries can be produced by a discrete set of coils or magnets that are easy to build.
 If there is no consideration of the coil and magnet complexity until after the plasma equilibrium has been chosen, complicated coils can result.
 It may be the case that small changes in the plasma geometry that result in small improvements in the plasma performance may also result in large disadvantageous changes in the coil geometry.
 Integrated algorithms that incorporate both plasma performance and coil complexity are therefore desirable \citep{Boozer2015,Landreman2016,Drevlak2019}.
 
 Measures of coil complexity can be included in the optimization cost function using the two-step method~\citep{Drevlak2019,Adjoint5}. This comes at the cost of computing the coil geometry or some approximation of it at every stage of the fixed-boundary plasma optimization. 

 Another approach for the combined plasma-coil design is the direct coil optimization using a free-boundary equilibrium code~\citep{Strickler_Integrated_2002,HudsonNCSX}.
 The direct coil optimization approach avoids the need for the inverse coil code because the coil geometry is taken to be the independent degree of freedom in the optimization, i.e., the coil geometry is assumed to be known.
 The external magnetic field that is required as input by the free-boundary equilibrium calculation is computed using the Biot-Savart formula.
 Metrics of coil complexity can easily be included in the optimization cost function. 

 There are direct coil optimization approaches that dispense with the equilibrium calculation and instead optimize the properties of the vacuum field, which may be thought of the trivial (zero plasma-current, zero pressure) plasma equilibrium state.
 This should be sufficient for designs that do not target high plasma pressure; however, finite-$\beta$ effects are important: the pressure-induced Shafranov shift \citep{ShafranovShift} can affect stability and confinement.
 A related direct coil optimization uses a near-axis analytic approximation to the equilibrium state \citep{Giuliani2020}, which might be sufficient for large aspect-ratio configurations but perhaps not so for ``compact'' stellarators.
Also, recent work has exploited analytical gradient information and adjoint methods for a variety of optimization problems \citep{Zhu_2018,Adjoint1,Adjoint2,Adjoint3,Adjoint4,Adjoint5}.
    
 In this paper, we review and categorize different combined plasma-coil optimization.
 We discuss free-boundary calculations and propose an alternative method to calculate the virtual-casing self-consistent vacuum magnetic field,
 which reduces the cost of a free-boundary calculation to that of a fixed-boundary calculation.
 In \Sec{threefunctionals}, the multi-region relaxed magnetohydrodynamic (MRxMHD) energy functional, the coil-penalty functional and the virtual casing integral are described. 
 In \Sec{functionalvariations}, we present the relevant variational calculus for the free-boundary equilibrium calculation, the coil geometry optimization and the virtual casing integral that are employed in the combined plasma-coil optimization algorithms discussed in \Sec{optimization}. 
 In \Sec{galerkin}, we discuss the construction of the magnetic field using a (weak) Galerkin method.
 In \Sec{optimization}, we examine various fixed- and free-boundary algorithms for the combined plasma-coil optimization that differ in which quantity, which we denote with $\bmz$, is chosen to be the independent degree-of-freedom, and derive the derivatives of the plasma equilibrium and the coil geometry with respect to $\bmz$ in terms of derivatives of the plasma-energy functional, the coil-penalty functional, and the virtual-casing integral.\footnote{If the geometry of the coils is chosen to be the independent degree-of-freedom, then the Biot-Savart integral, \Eqn{biotsavart}, is required instead of the coil-penalty functional.}
 Finally, in \Sec{discussion}, we discuss the results of this paper.

 \section{Plasma equilibria and supporting coils}
 \label{sec:threefunctionals}
 
 This paper builds upon the construction of the plasma equilibrium as a stationary point of the multi-region relaxed magnetohydrodynamic (MRxMHD) energy functional, which we now describe.%
 \footnote{We have chosen to use the MRxMHD functional and not the ideal MHD functional because the ideal MHD functional encounters problems near rational surfaces. However it is straightforward to extend the work of this paper to use ideal MHD. In fact, we believe that most of the results of this paper can be extrapolated to ideal MHD.}

 The \emph{computational domain}, $\Omega\subset \mathbb{R}^3$, is considered to be a solid torus with \emph{computational boundary} $\cD \equiv \partial \Omega$.
 The latter is a prescribed toroidal surface, which unless explicity stated otherwise is held fixed throughout the calculation.
 The computational domain contains the \emph{plasma volume}, $\cV\subset \Omega$, a smaller solid torus enclosed by the \emph{plasma boundary} $\pcV\equiv \partial {\cal V}$.
 The plasma volume is further partitioned into $v = 1, \dots, N_V$ nested toroidal sub-volumes, $\cV_v$, which are separated by a set of \emph{ideal interfaces}, ${\cal I}_v$ so that $\partial \mathcal{V}_v=\cI_v\cup \cI_{v-1}$, with the outermost interface coinciding with the plasma boundary, $\cI_{N_V} = \pcV$.
 The innermost sub-region is a simple (solid) torus, and each other sub-region is a toroidal annulus which may be thought of as a ``hollow'' torus. 
 In each $\cV_v$, the magnetic field is written as the curl of the magnetic vector potential, $\bmB_v = \nabla \times {\bm{A}}_v$. 
 In the innermost toroidal region, the toroidal flux, computed as a poloidal loop integral of ${\bm{A}}_1$ on $\cI_1$, is constrained to match a prescribed value. 
 In each annular region, both the enclosed toroidal and poloidal fluxes, computed respectively as the difference between the poloidal and toroidal loop integrals of ${\bm{A}}_v$ on $\cI_v$ and $\cI_{v-1}$, are constrained. 
 In each region, the magnetic helicity, defined as the volume integral of ${\bm{A}}_v \cdot \bmB_v$, is also constrained. 
 The magnetic field in each region is taken to be tangential to that region's boundary, $\bn\cdot \bm{B}_v=0$ on $\cI_v$ where $\bn$ is a unit outward normal vector, but otherwise the topology of the fieldlines is unconstrained and magnetic islands and irregular field lines are allowed in each ${\cal V}_v$.
 
 For free-boundary calculations, by which we mean that the plasma boundary is allowed to move, we include the \emph{vacuum} region $\cV_+=\Omega\setminus \cV$, with the inner boundary of $\cV_+$ coinciding with $\pcV$ and its outer boundary with $\cD$.
 It is on $\cD$ that the normal plasma field plus the normal external field is required to equal the normal total field. 
 
 By construction, there are no currents inside $\cV_+$. 
 We may employ a scalar potential and write the \emph{vacuum field} as $\bmB_{+} = \nabla \Phi$.\footnote{We may also use the magnetic vector potential to describe the magnetic field in $\cV_+$, and this can have advantages \citep{Hudson2020}, but we will not discuss this possibility here.}
 To obtain a unique solution, constraints are imposed on the linking and net toroidal plasma currents, which are computed as loop integrals of $\bmB_+$. We take $\bmB_+$ to be tangential to the plasma surface, $\bm{n} \cdot \nabla \Phi|_{\pcV}=0$ where $\bm{n}$ is a unit outward normal vector on $\pcV$. 
 The normal field on the computational boundary $\cD$, $B_{T,n} \equiv \bar{\bm{n}} \cdot \nabla \Phi|_{\cD}$ where $\bar{\bm{n}}$ is normal to $\cD$, is the sum, $B_{P,n} + B_{E,n}$, of the plasma field and the external field, neither of which are necessarily known {\it a priori}.
 Generally, $B_{T,n} \ne 0$.
  
 The equilibria that we consider are stationary points of the following energy functional,
 \begin{eqnarray}
 \calF & \equiv & 
 \sum_{v=1}^{N_V} \calF_v + \calF_+,
 \label{eqn:MRxMHDBT}
 \end{eqnarray}
 where 
 \begin{eqnarray}
 \calF_v & \equiv & 
 \int_{{\cal{V}}_v} \left( \frac{p_v}{\gamma-1} + \frac{B_v^2}{2} \right) \diff v 
 -
 \frac{\mu_v}{2} \left[ \int_{{\cal{V}}_v} {\bm{A}}_v \cdot \bmB_v \diff v - H_{v} \right], 
  \label{eqn:Fv}
  \\
 \calF_+ & \equiv & 
 \int_{\cV_+} \frac{1}{2} \nabla \Phi \cdot \nabla \Phi \, \diff v 
 -\int_{\cD} \bar \Phi \bm{B}_T  \cdot \diff \bar\bs,
  \label{eqn:F+}
 \end{eqnarray}
 with $p_v$ the pressure in the $v$-th region\footnote{Here, the pressure is understood in units of $\mu_0 \times[J/m^3]$ where $\mu_0=4\pi \times 10^{-7}[H/m]$ is the vacuum permeability}, $\gamma$ is the specific heat ratio (adiabatic index), and $H_v$ is the prescribed value of the magnetic helicity in $\calV_v$.
 The constraints on the toriodal and poloidal fluxes in the plasma volumes, $\calV_v$, are implied; that is, the fields $\bmB_v$ that we consider in \Eqn{Fv} are restricted to the class of divergence-free vector fields that are tangential to the interfaces and have the prescribed values of the toroidal and poloidal fluxes.
 The helicity constraints are enforced explicitly using the Lagrange multipliers $\mu_v$.
 The loop integral constraints on $\nabla\Phi$ are also implied: the allowed $\Phi$ in \Eqn{F+} is restricted to the class of multi-valued scalar functions with prescribed loop integrals of $\nabla\Phi$.
 We write $\bar\Phi\equiv\Phi|_{\cD}$ to denote the scalar potential evaluated on $\cD$ to distinguish it from $\Phi\equiv\Phi|_{\pcV}$ on $\pcV$. 
 Similarly, we write $\diff\bar\bs$ to denote an infinitesimal surface element on $\cD$.
 
 We may call $\calF$ the \emph{free-plasma-boundary fixed-computational-boundary generalized-boundary-condition MRxMHD energy functional}.
 This is to accommodate the common understanding in the magnetic confinement community that a free-boundary equilibrium calculation allows the plasma boundary to move, which it does in this case, and the common terminology in the broader mathematical community that refers to the ``boundary'' as boundary of the computational domain, which is fixed in this case.
 We refer to the boundary conditions as ``generalized'' because the total normal field, $B_{T,n}$ is not constrained to be zero.
 For brevity, we hereafter refer to $\calF$ simply as the energy functional.
  
 Using the virtual-casing principle \citep{VirtualCasing}, the normal plasma field, $B_{P,n}$, produced by currents within the plasma volume is computed on the computational boundary as
 \begin{eqnarray}
 B_{P,n}(\bar\bmx) & \equiv & \left( - \frac{1}{4\pi} \int_{\pcV} 
 \frac{\left( \bmB_T|_{\pcV_+} \times \diff \bs \right) \times \br}{r^3} \right) \cdot \bar\bn(\bar\bx),
 \label{eqn:BPn}
 \end{eqnarray}
 where ${\bm{r}} \equiv \bmx - \bar \bmx$ for $\bmx \in \pcV$ and $\bar\bmx\in\cD$, and we use $\diff \bs$ to denote a surface element of $\pcV$.
 Here, $\bmB_T|_{\pcV_+}=\nabla \Phi|_{\pcV_+}$ is the total magnetic field immediately outside of $\pcV$.
 To accommodate for the possible existence of sheet currents lying {\em on} the plasma boundary, we must use in \Eqn{BPn} the total magnetic field lying immediately {\em outside} the plasma boundary; that is, we must use the magnetic field on the inner boundary of the vacuum region. 
 When that information is not available, we can use the magnetic field immediately inside the plasma boundary.
 In \Sec{galerkin}, we shall exploit the circumstance that the evaluation of the plasma field on the computational boundary is a non-local, linear operator of the scalar potential, $\Phi$.
 
 The difference between the total normal magnetic field and the plasma normal field on $\cD$, which we call the {\it required} external normal field, must be provided by an external source. A plasma cannot create its own magnetic bottle \citep{virial}.
 This quantity, $D_n \equiv B_{T,n}-B_{P,n}$, is very closely related to $B_{E,n}$, the {\it provided} external field, but they may differ. The difference is what we later call the ``coil error''.
 
 For clarity of exposition regarding the externally applied field and to facilitate discussion of the various combined plasma-coil optimization algorithms considered in \Sec{optimization}, we imagine a typical and easily differentiable example; namely, that the external magnetic field is provided by a set of $i = 1, \dots, N_C$ current-carrying one-dimensional curves (filaments) with geometry $\bmc_i$ that produce a magnetic field given by the Biot-Savart law,
 \begin{eqnarray} \bmB_E(\bar \bmx) & = &
 -\sum_i \frac{\mu_0 I_i}{4\pi}\int_{\bmc_i} \frac{\br \times \diff {\bm{l}}}{r^3}
 \label{eqn:biotsavart}
 \end{eqnarray}
 where $I_i$ is the current in the $i$-th coil and $\diff {\bm{l}}$ is the differential line segment.
 For brevity, we shall hereafter ignore the dependency on the magnitude of the currents and we set $I_i = 4\pi/\mu_0$. 
 It is a simple matter to extend the following to allow for the variation of the coil currents.
 It is also a simple matter to extend the following to the case where a surface potential on a prescribed winding surface provides the external field \citep{NESCOIL,REGCOIL}, or to the case where a ``finite-build'' approximation is used for the coils \citep{Singh2020,McGreivy2020,Li2020},
  or when permanent magnets are included \citep{PermanentMagnets,Zhu2020,Landreman2020}. 
 
 If the required external field, $D_{n}$, is given, the geometry of the coils may be chosen to minimize a suitable ``coil-penalty'' functional; for example \citep{NESCOIL,REGCOIL,Zhu_2018,HUDSON2018},
 \begin{eqnarray}
 \calE & \equiv & \varphi_2 + \omega L, \label{eqn:coilpenaltyfunctional}
 \end{eqnarray}
 where $\varphi_2 \equiv \displaystyle \int_{\cD} \textstyle \frac{1}{2} Q^2 \diff\bar s$ is the commonly used quadratic-flux error functional, 
 $Q = D_n - B_{E,n}$ where $B_{E,n}[\bmc,\cD]$ is the normal component of $\bmB_E$ on $\cD$ produced by the filamentary coils, where we use $\bmc$ to denote the geometry of all the coils; i.e., $\bmc \equiv \{ \bmc_i, i = 1, \dots N_C \}$.
 We include a regularization term, $L = L[\bmc]$, which we take to be the total length of the coils, and $\omega$ is a scalar penalty.
 
 The geometry of the optimal coils is given by $\delta \calE/\delta \bmc = 0$.
 This equation may be solved using descent algorithms \citep{FOCUS1,HUDSON2018} or Newton-style methods \citep{Zhu_2018}. 
 Practically, it is not possible to obtain a coil set that exactly produces the required normal field and generally $\varphi_2\ne 0$.
 This is why the required external normal field, $D_n$, must be distinguished from the actual external normal field, $B_{E,n}$. 
 Upon solving $\delta \calE/ \delta \bmc = 0$, $\varphi_2$ measures what we call the {\it coil error}.\footnote{Note: this is not a ``numerical'' error, in the sense that this error will {\it not} decrease as the numerical resolution of the discrete set of coils increases. The coil error results from the inevitable requirement that the external field is provided by a discrete set of coils (or permanent magnets \citep{PermanentMagnets,Zhu2020,Landreman2020}) which also are constraint by other (engineering) properties.}
 
 In the above, we have described the three basic components of the combined plasma-coil optimization algorithms that we consider in \Sec{optimization}; namely, the energy functional, $\calF$, the Biot-Savart law, the coil-penalty functional, $\calE$, and the evaluation of the normal plasma field, $B_{P,n}$, on $\cD$ using the virtual-casing functional.
 In the following section, \Sec{functionalvariations}, we present the first and second variations of $\calF$, $\calE$ and $B_{P,n}$ that will be required in \Sec{optimization}.
 In \Sec{galerkin}, we elaborate upon the numerical construction of the magnetic fields and show that, instead of prescribing the total normal magnetic field, $B_{T,n}$, on $\cD$, computing the equilibrium and then determining the coil geometry that provides (approximately) the required external normal field, we may directly consider the case where the external normal field, $B_{E,n}$, on $\cD$ is given.
 We can solve for the {\it virtual-casing self-consistent} vacuum magnetic field using a Galerkin method.
 In \Sec{optimization}, we show that the derivatives described in \Sec{functionalvariations} and the Galerkin construction of the magnetic fields may be combined to construct a variety of combined plasma-coil stellarator optimization algorithms.
 
 \section{Variation of the energy, coil-penalty and virtual-casing functionals} 
 \label{sec:functionalvariations}

 In this section, we present the variations of the energy, coil-penalty and virtual casing functionals  
 that are used in \Sec{optimization}.
 The variational calculus provides useful insight into the physics and mathematics of the different optimization problems, as we shall comment upon in \Sec{discussion}.

 Requesting that the first variations of the energy functionals vanish results in weak formulations of the various coupled boundary value problems. 
 When the boundary data is ``smooth" enough, it is expected that weak solutions coincide with (strong) solutions. 
 In what follows, we assume the fields are sufficiently regular to apply integration by parts and derive local Euler-Lagrange equations. 
 Conversely, we convert any linear elliptic partial differential equations subject to boundary conditions into variational problems (weak formulation) simply by integrating against arbitrary variations. 
 In the case where the bilinear functional thus obtained is symmetric (self-adjoint), the variational problem can be associated to an energy minimization principle. However, as we will see in \Sec{freeboundary}, the weak formulation of certain boundary value problems do not necessarily result in self-ajoint bilinear functionals, in which case there is no immediate energy minimization principle.

 The weak formulation is the basis of our numerical representation of solutions, in particular leading to a spectral-Galerkin method where our fields are approximated by a finite linear combination of Fourier modes and orthogonal polynomials. The coefficients of the linear combination become our degrees of freedom and the weak formulation boils down to a matrix inversion.
 
 \subsection{Energy functional}
 \label{sec:energyfunctionalvariations}

 In the plasma volumes, we write $\bb_v = \curl \ba_v$, and thus $\delta \bb_v = \curl\delta\ba_v$. 
 We use that the fact that the fields in the vicinity of the interfaces obey ideal MHD, so that the variation of the magnetic field on the interfaces can be expressed with an displacement vector $\bxi_v$, $\delta \bb_v = \curl(\bxi_v\times\bb_v)$, i.e. $\delta \ba_v = \bxi_v\times \bb_v + \nabla g_v$, where $g_v$ is a  gauge potential.
 In MRxMHD, the mass is constrained in each volume ${\cal V}_v$ by the isentopic ideal-gas constraint, $p_v = a_v / V_v^{\gamma}$, where $V_v$ is the volume of the $v$-th region, $V_v = \displaystyle \int_{{\cal V}_v} \diff v$, and $a_v$ is a constant. 
 The first variation with respect to deformations of the volume boundary is given by $\displaystyle \delta p_v = - \gamma p_v \left(\displaystyle\int_{{\cI}_v} \bxi_v\cdot\diff \bm{s}-\int_{{\cI}_{v-1}} \bxi_{v-1}\cdot\diff \bm{s}\right)/V_v$.
 
 Using $\bb_v \cdot {\bn}_v = 0$ on the interfaces, where $\bn_v$ is the normal vector of the $v$-th interface, one can derive
 \begin{eqnarray}
 \delta {\cal F}
 & = & \sum_{v=1}^{N_v}\left(\int_{{\cal{V}}_v} \frac{\delta{\cal F}_v}{\delta \ba_v} \cdot \delta \ba_v \diff v  + \int_{\cI_v} \frac{\delta {\cal F}_v}{\delta \bxi_{v}} \cdot \bxi_{v} \diff s\right) + \delta \mathcal{F}_+, 
 \end{eqnarray}
 where
 \begin{eqnarray}
 \frac{\delta{\cal F}_v}{\delta \ba_v} & = & \curl\bb_v - \mu_v \bb_v, \label{eqn:Beltrami} \\
 \frac{\delta{\cal F}_v}{\delta \bxi_{v}} & = & - \biggl[ \! \biggl[ p_v + \frac{B_v^2}{2} \biggl] \! \biggl] \bn_v.
 \label{eqn:forcebalance}
 \end{eqnarray}
 In vacuum, the first variation in $\calF_+$ with respect to variations in $\Phi$ is
 \begin{eqnarray}
 \delta \calF_+ & = &
	\int_{{\cal V}} \frac{\delta \calF_+}{\delta \Phi} \delta \Phi \, \diff v 
+ 	\int_{\cD}\delta \bar\Phi\,  \left( \nabla \bar \Phi - \bmB_T \right) \cdot \diff\bar{\bm{s}} - \int_{\pcV} \delta \Phi \,\nabla  \Phi \cdot \diff \bm{s}.
 \label{eqn:firstvariationphi}
 \end{eqnarray}
 Setting the functional derivative,  
 \begin{eqnarray}
 \frac{\delta \calF_+}{\delta \Phi} & = & -\nabla \cdot \nabla \Phi,
 \label{eqn:Laplace}
 \end{eqnarray}
 to zero leads to the Laplace equation.
 The other terms provide corresponding Neumann boundary conditions.

 The second variation with respect to deformations of the interface geometry of ${\cal{F}}_v$ is given by 
 \begin{eqnarray}
 \delta^2 {\cal{F}}_v & = & \int_{{\cal{V}}_v} \delta \ba_v \cdot ( \curl\curl\delta \tilde{\ba}_v-\mu \curl \delta \tilde{\ba}_v )\diff v  
 + \gamma p_v \frac{\int_{\cI_v} \bxi_v\cdot \diff \bm{s} \int_{\cI_v} \tilde \bxi_v\cdot \diff \bm{s}}{V_v}\nonumber\\
	 &-&  \int_{\cI_v}  ( \bb_v\cdot\curl \delta\tilde\ba)\bxi_v  \cdot\diff\bm{s} 
	 \nonumber
	 -  \int_{\cI_v} \delta \ba_v \cdot(\mu_v \bb_v - \nabla \times \bb_v)\tilde{\bxi}_v  \cdot\diff\bm{s} \nonumber \\
	 &-&  \int_{\cI_v} \tilde{\bxi}_v\cdot\bn_v \nabla\cdot\left((p_v+\frac{B_v^2}{2})\bxi_v\right) \diff s,\label{eqn:ddFdAdxi}
	\end{eqnarray}
	where the tilde ($\sim$) indicates that the second variation can differ from the first. Here, for simplicity, we allowed only one of the two interfaces to vary. For a complete expression (see \Eqn{deltadeltaF}), one has to be careful since cross-terms, including integrals over two different interfaces, appear in the $\gamma p_v$ term. 
	 
 In vacuum, the second variation with respect to the scalar potential is given by:
 \begin{eqnarray}
	\delta^2\calF_+ = -\int_{\cal V} \delta \Phi \, \nabla\cdot\nabla \delta \tilde{\Phi} \, \diff v - \int_{\MB} \delta \bar\Phi \, \bar\bn\cdot\nabla \delta \tilde{\bar\Phi} \, \diff \bar s.
	\label{eqn:secondvariationA}
	\end{eqnarray}	
 Collecting all the different terms for the second variation, we obtain
\begin{eqnarray}
\delta^2 {\cal F} &=& \sum_v\bigg(\int_{{\cal{V}}_v} \delta \ba_v \cdot ( \curl\curl\delta \tilde{\ba}_v-\mu_v \curl \delta \tilde{\ba}_v )\diff v  \nonumber \\
 &+&\gamma\left( \frac{p_v}{V_v}+\frac{p_{v+1}}{V_{v+1}}\right)
 \left(\int_{\cI_v} \bxi_v\cdot \diff \bm{s}\right)
 \left(\int_{\cI_v} \tilde \bxi_v\cdot \diff \bm{s} \right)\nonumber \\
 &-&\gamma \frac{p_v}{V_v}  
 \left[\left(\int_{\cI_{v-1}}\tilde\bxi_{v-1} \cdot \diff \bs \right)
 \left(\int_{\cI_v}\bxi_v\cdot \diff \bs\right) 
 + 
 \left(\int_{\cI_{v-1}}\bxi_{v-1}\cdot \diff \bs \right) 
 \left(\int_{\cI_v}\tilde\bxi_v\cdot \diff \bs\right)\right]\nonumber\\
	 &+&  \int_{\cI_v}  \biggl[ \! \biggl[  \bb \cdot ((\bb\times\nabla)\times\bn) \biggl] \! \biggl] \tilde\bxi_v \cdot \bn \,\,\bxi_v \cdot\diff\bm{s} \bigg) + \delta^2 \calF_+, \label{eqn:deltadeltaF}
 \end{eqnarray}
 where we used the equilibrium expressions; we included variations of both interfaces; we used $\bb \cdot ((\bb\times\nabla)\times\bn)=\bb^2 \nabla\cdot\bn - \bb\cdot\nabla \bn\cdot\bb$; and we set $\bxi_v$ and $\tilde \bxi_v$ proportional to the normal vector $\bn$ since only normal variations of the interfaces affect the plasma energy.
 The second variation of the energy functional determines the MRxMHD stability of the equilibrium state \citep{Kumar2020}.
 \footnote{For ideal MHD the first and second variations of the energy functional are well known \citep{Freidberg_2014}. 
 Codes like TERPSICHORE and CAS3D \citep{Anderson1990,CAS3D} can compute the second variation of the ideal MHD functional.}

 \subsection{Coil-penalty functional} 
 \label{sec:coilvariation}
	
 The coil-penalty functional, $\calE$ defined in \Eqn{coilpenaltyfunctional} is considered to be a functional of the coil geometry, $\bmc$, the required normal magnetic field, $D_n$, on the computational boundary, $\cD$, and on $\cD$ itself, which is however kept fixed here; i.e., $\calE=\calE[\bmc,D_n]$. 
 We may formally write the first variation, $\delta \calE$, with respect to variations $\delta\bmc$, and $\delta D_n$ as
 \begin{eqnarray}
 \delta \calE & = & \sum_{i} \int_0^{2\pi} \frac{\delta \calE}{\delta \bmc_i} \cdot \delta \bmc_i \, \diff \alpha 
 + \int_{\cD} \frac{\delta \calE}{\delta D_n} \cdot \delta D_n \, \diff \bar s,
 \end{eqnarray}
 where $\alpha$ is an angle-like curve parameter, $\bmc_i(\alpha+2\pi)=\bmc_i(\alpha)$.
 The functional derivatives are 
 \begin{eqnarray}
 \frac{\delta \calE}{\delta \bmc_i}  & = & \bmc_i' \times \left( \int_{\MB} {\bm{R}}_{i,n} Q \diff s + \omega \, \bkappa_i \right), \label{eqn:dEdc} \\
 \frac{\delta \calE}{\delta D_n}  & = & Q, \label{eqn:dEdDn}  
 \end{eqnarray}
 where $\bmc_i^\prime$ is the derivative of the $i$-th coil with respect to $\alpha$. We have written $\bR_i \equiv 3 \, \br_i \br_i / r_i^5 - \bi / r_i^3$ and defined ${\bm{R}}_{i,n} = \bR_i \cdot \bm{n}$, where $\br_i$ is the displacement between a point on the $i$-th coil and the evaluation point on $\cD$, and  $\bi$ is the identity tensor or \emph{idemfactor}. 
 The curvature of the $i$-th coil is $\bkappa_i \equiv \bmc_i' \times \bmc_i'' / \bmc_i'^3$ and the functional derivatives are generalized expressions of the ones presented in \citep{Dewar1994,HUDSON2018}. 
 We used the expression for the variation of the normal component of the magnetic field with respect to the coil geometry
 \begin{eqnarray}
 \delta B_{n} = \oint_i \frac{\delta B_n}{\delta \bmc_i} \cdot \delta\bmc_i  \diff{\alpha_i}, \label{eqn}
 \end{eqnarray}
 with the functional derivatives
  \begin{eqnarray}
\frac{\delta B_n}{\delta \bmc_i} =\bmc_i' \times {\bm R}_{i,n}. \label{eqn:dBndc}
 \end{eqnarray}
 
 The required second variations of $\calE$ are \citep{HUDSON2018} 
 \begin{eqnarray}
 \delta^2 \calE & = & \sum_{i,j} \oint_i \oint_j \delta \bmc_i \cdot \frac{\delta^2 \calE}{\delta \bmc_i \delta\bmc_j} \cdot \delta \bmc_j \diff{\alpha_i} \diff{\alpha_j}, \\
 \delta^2 \calE & = & \sum_i \oint_i \diff{\alpha} \oint_{\MB} \diff \bar s \; \delta \bmc_i \cdot \frac{\delta^2 \calE}{\delta \bmc_i \delta D_n} \, \delta D_n,
 \label{eqn:E2}
 \end{eqnarray}
 where
 \begin{eqnarray}
 \frac{\delta^2 \calE}{\delta \bmc_i\delta\bmc_j} 
 & = & 
 \oint_{\MB} (\bmc_i' \times {\bm{R}}_{i,n}) (\bmc_j' \times {\bm{R}}_{j,n}) \diff \bar s, \;\; \mbox{\rm (for $i \ne j$)}, \label{eqn:ddEdcdc} \\
 \frac{\delta^2 \calE}{\delta\bmc_i \delta D_n } 
 & = & 
 \bmc_i' \times  {\bm{R}}_{i,n}. \label{eqn:ddEdcdDn} 
 \end{eqnarray}
 For the case where $i=j$ the variation cannot be written in such a compact form but also does not provide much insight.
 The other second variations of $\calE$ are not required in \Sec{optimization} and are omitted for brevity.
 
 We only need the variation with respect to the geometry of the computational boundary, $\MB$, when $D_n=-B_{P,n}=-\bb_{P}\cdot\bar\bn$. 
 Since there is an explicit dependence with respect to $\bar \bn$, we calculate the combined variation of $\calE$ and $B_{P,n}$ with respect to variations in $\MB$, which allows us to express the results in compact form. 
 The first variation is
 \begin{eqnarray}
 \delta \calE & = &  \int_{\cD} \frac{\delta \calE}{\delta \cD} \cdot \delta \cD \; \diff \bar s,
 \end{eqnarray}
 where
 \begin{eqnarray}
 \frac{\delta \calE}{\delta \cD} & = & \left( -\textstyle \frac{1}{2}Q^2 \nabla \cdot \bar \bn - \left( \bb_{P,s} + \bb_{E,s} \right) \cdot\nabla Q \right) \bar \bn,  \label{eqn:dEdD} 
 \end{eqnarray}
 where we used Eqn. 12 from \citep{Dewar1994}.
 The notation $\bm{f}_{s} = ( \bi - \bar \bn \bar \bn ) \cdot \bm{f}$ denotes the surface projection of an arbitrary vector or tensor, $\bm{f}$, onto a surface, which in this case is the computational boundary.
 The second variation of $\calE$ with respect to $\MB$ and $\bmc_i$ is
 \begin{eqnarray}
 \delta^2 \calE & = & \sum_i \oint_{\bmc_i} \delta \bmc_i \cdot \oint_{\cD} \frac{\delta^2 \calE}{\delta \bmc_i \delta \cD} \cdot \delta \cD \; \diff \bar s \, \diff\alpha,
 \end{eqnarray}
 where
 \begin{eqnarray}
 \frac{\delta^2 \calE}{\delta \bmc_i \delta \cD} 
 & = & 
 \bmc_i^\prime \times \left( Q \bR_i \cdot \bm{H} - {\bR}_{i,s} \cdot \nabla {\cal H} + (\bmB_{P,s}+\bmB_{E,s}) \cdot \nabla {\bm{R}}_{i,n}  \right) \bar{\bm{n}}, \label{eqn:ddEdcdD}
 \end{eqnarray}
 where the mean curvature of $\cD$ is $\bm{H} \equiv - \bar{\bm{n}} \, \left( \nabla \cdot \bar{\bm{n}} \right)$. \Eqn{ddEdcdD} is a more general form of Eqn.~12 of~\citep{HUDSON2018} where $D_n=0$.
 
 \subsection{Plasma normal field}
 \label{sec:BPnvariations}
 
 The normal plasma field, $B_{P,n}$, on $\cD$, as defined in \Eqn{BPn}, is considered to be a functional of the plasma boundary, ${\cal S}$, the total (tangential) magnetic field, $\bmB_T|_{\pcV}$ on $\pcV$, and on $\cD$, which is kept fixed; i.e., $B_{P,n}=B_{P,n}[{\cal S},\bmB_T|_{\pcV}]$. 
 The normal component of the magnetic field produced by plasma currents, $B_{P,n}$, given in \Eqn{BPn} may be written as
 \begin{eqnarray}
 B_{P,n} & = & - \frac{1}{4\pi} \int_{\pcV} \bmB_{T}|_{\pcV} \cdot \bMt \cdot \diff\bm{s},
 \end{eqnarray}
 where the anti-symmetric tensor, $\bMt \equiv ( {\bm{r}} \, \bar{\bm{n}} - \bar{\bm{n}} \, {\bm{r}} ) / r^3$, is defined, where ${\bm{r}} = \bmx-\bar\bmx$ for $\bmx \in \pcV$ and $\bar\bmx \in \cD$.
 The first variation setting $\delta \MB =0$ is
 \begin{eqnarray}
 \delta B_{P,n} & = & 
 \int_{\pcV} \frac{\delta B_{P,n}}{\delta \pcV} \cdot \delta \pcV \; \diff  s 
 + \int_{\pcV} \frac{\delta B_{P,n}}{\delta \bmB_T|_{\pcV} } \cdot \delta \bmB_T|_{\pcV} \; \diff s, 
 \end{eqnarray}
 where
 \begin{eqnarray}
 \frac{\delta B_{P,n}}{\delta \pcV} & = & - \frac{1}{4\pi} \nabla \cdot ( \bmB_T|_{\pcV} \cdot \bMt ), \label{eqn:dBPndS} \\
 \frac{\delta B_{P,n}}{\delta \bmB_T|_{\pcV}} & = & - \frac{1}{4\pi} \bMt \cdot \bn, \label{eqn:dBPndBT}
 \end{eqnarray}
 where we used again Eq. 12 from \citep{Dewar1994} for the first expression. 
 Note that only $\bMt$ and $\bR$ depend on $\bar \bx$.
 The relevant variation with respect to the computational boundary $\MB$ is presented in the previous section.
 The second variations are not required in the following.

 \subsection{Supplied external field}
 \label{sec:freeboundary}
 
 To enable efficient free-boundary optimization algorithms, which we describe in \Sec{optimization}, we revisit the energy functional; in particular, we pay attention to the boundary condition for the normal field on $\cD$.
 
 It is perfectly legitimate to prescribe the {\it total} normal field, $B_{T,n}$, on $\cD$.
 Indeed, every fixed-boundary equilibrium calculation does as much.
 As described in \Sec{threefunctionals}, the total normal field is the sum of the plasma normal field and the external normal field, $B_{T,n} = B_{P,n} + B_{E,n}$, neither of which are known {\it a priori} in fixed-boundary calculation.
 
 It seems unlikely that there are any circumstances for which we will {\it a priori} know $B_{P,n}$, except for the trivial equilibrium, the vacuum, for which $B_{P,n}=0$.
 After all, this is why the equilibrium calculation is required, to determine the plasma currents and the magnetic field that they produce. 
 
 In contrast, a particularly important calculation arises when $B_{E,n}$ is known.
 In this case, if we were to proceed with the approach of specifying $B_{T,n}$ on $\cD$, then some type of ``self-consistent'' iteration, for example, must be implemented to determine the $B_{T,n}$ that satisfies the {\it matching condition}, namely that $B_{T,n} - B_{P,n}[\bb_T|_\mathcal{S}] = B_{E,n}$, where $B_{P,n}[\bb_T|_{\pcV}]$ may be considered to be a linear, nonlocal operator acting on the tangential total field on the plasma boundary $\bb_T|_{\pcV}$, which is only known after the equilibrium has been computed.
 In the current implementation of free-boundary SPEC \citep{Hudson2020}, for example, a Picard iteration over $B_{T,n}$ is required.
 
 Here, we present a more direct strategy. 
 The boundary value problem in the vacuum region is to find $\bb_+ = \nabla\Phi$ such that $\nabla \cdot \bb_+ = \nabla \cdot \nabla \Phi = 0$ on $\cV_+$ satisfying boundary conditions $\bn\cdot\nabla\Phi=0$ on $\pcV$ and $\bar\bn\cdot \nabla\bar\Phi - B_{P,n}[\nabla\Phi|_{\pcV}]= B_{E,n}$ on $\cD$. 
 This translates into the weak problem of finding $\Phi$ such that
 \begin{align}
 \label{eqn:vp}
     \int_{\cV_+} \!\! \nabla\psi \cdot \nabla\Phi \diff v 
     + \int\limits_{\cD} \! \int\limits_{\pcV} \frac{(\diff \bar{\bs}\times \nabla\bar\psi)\cdot(\diff \bs\times \nabla\Phi)}{|\bar\bx-\bx|}   = \int_{\cD} \bar\psi \, B_{E,n} \diff \bar s, \quad \forall \psi.
 \end{align}
where $\psi$ is an arbitrary (test) function. An important solvability condition is that the normal field of the coils is consistent with a divergence-free field, $\displaystyle \int_{\cD} B_{E,n} \diff \bar s =0 $.
 
 As it stands, it is not possible to cast this weak problem into an energy minimization problem because the second term on the left-hand side of \Eqn{vp} spoils the required symmetry of the functional. The first variation of a quadratic energy functional necessarily results in symmetric bi-linear forms. 
 
 Energy principles do exist for exterior Neumann problems in the form of boundary integral methods \citep{Giroire1978}. 
 Their formulation however requires the plasma boundary and the computational boundary to coincide. 
 The NESTOR code is a good example~\citep{MERKEL1986}, as well as BIEST~\citep{BIEST}. There are several disadvantages and challenges in implementing these energy principles. First, as seen in \Eqn{vp} in the limit $\cD\to\pcV$, the integral operators tend to have singular kernels, which requires delicate numerical treatment. 
 Second, in free-boundary calculations, the plasma boundary is varied to achieve force-balance.
 This requires updating the numerical representation of the integral operators as well as the source term on the right-hand side of \Eqn{vp} every time the boundary changes. 
 For this, the external field (from coils) is effectively needed in an entire volume not just on a single surface, which comes with a sizeable computational cost~\footnote{Readers familiar with VMEC will recognize this as the task of computing the magnetic field  on a Cartesian grid from coils via Biot-Savart law, known as \emph{mgrid} file.}.
 
 In the following section, \Sec{galerkin}, we present a free-boundary equilibrium approach for a given external field that is based on a weak (Galerkin) method for constructing the required magnetic fields while keeping the computational boundary fixed and separate from the plasma boundary so that only the normal component of the external field on the computational boundary need be evaluated and only once.
 The virtual-casing self-consistent vacuum field is obtained as the solution to a set of linear equations, given below in \Eqn{selfconsistentvacuumfield}.
 The matching condition is enforced directly in the computation of the vacuum field, and this removes the self-consistent iteration mentioned above that is otherwise required to match the equilibrium to the provided external field.
 
 \section{Galerkin method for constructing the vacuum field}
 \label{sec:galerkin}
 
 Galerkin methods for constructing weak solutions to partial differential equations are well-known.
 In the context of MRxMHD, a Galerkin method for constructing the magnetic fields for fixed- and free-boundary MRxMHD equilibria is described in \citep{Hudson2012,Hudson2020,Zhisong2020}.
 In this paper, we restrict our attention to the problem of constructing $\bmB_+$.
 
 A standard numerical approach for finding stationary points of $\calF_+$ is to represent the scalar potential using a mixed Chebyshev-Fourier representation; e.g., $\Phi(s,\theta,\zeta) = I \, \theta + G \, \zeta + \sum_{l,m,n} \Phi_{l,m,n} T_l(s)\exp(im\theta-in\zeta)$, where $(s,\theta,\zeta)$ is a suitable toroidal coordinate system and $T_l(s)$ is the $l$-th Chebyshev polynomial, and $I$ and $G$ are given by the prescribed value of the loop integrals$/2\pi$.
 When this representation is inserted into \Eqn{Fv}, and assuming that $\pcV$ and $\MB$ do not change, the problem of calculating the vacuum field amounts to solving
 \begin{eqnarray}
 \cA \cdot \bphi = \bmB_T,
 \label{eqn:Laplacematrix}
 \end{eqnarray}
 where $\bphi$ represents the vector of independent degrees-of-freedom, namely the $\Phi_{l,m,n}$ and the Lagrange multipliers that may be used to enforce the constraints, and $\cA$ is a symmetric matrix whose elements are the second derivatives of $\cF_{+}$ with respect to these degrees-of-freedom and involve volume integrals of coordinate metric elements and the Chebyshev-Fourier functions.
 The matrix $\cA$ depends on the geometry of $\pcV$ and $\cD$; i.e., $\cA=\cA[\pcV,\cD]$.
 The right-hand-side vector, $\bmB_T$, contains the Fourier harmonics of $B_{T,n}$.
 The system of linear equations given in \Eqn{Laplacematrix} is essentially a discrete realization of Laplace's equation, $\nabla \cdot \nabla \Phi = 0$, with the supplied boundary conditions and loop integrals, written in matrix form.
 We can determine how the vacuum magnetic field varies with $\pcV$, ${\cal D}$ and $B_{T,n}$ using matrix perturbation methods,
 \begin{eqnarray}
 \cA \cdot \delta \bphi & = & - \delta \cA \cdot \bphi + \delta \bmB_T,
 \label{eqn:vacuumfieldderivative}
 \end{eqnarray}
 where $\delta {\cal A} = \partial \cA / \partial \pcV \cdot \delta \pcV + \partial {\cal A} / \partial \MB \cdot \delta \MB$.
 
 From \Eqn{BPn}, we recognize that the magnetic field produced by the plasma currents is a linear functional of the total (tangential) magnetic field on the immediate outside of the plasma boundary, namely $\bmB_+|_{\pcV_+}$.
 The immediate outside of $\pcV$ is the inner boundary of the vacuum region, $\cV_+$, and the magnetic field in $\cV_+$, namely $\bmB_+$, is the gradient of the scalar potential.
 The Fourier harmonics, $B_{P,n}$, of the plasma field on $\cD$ are linear in the degrees-of-freedom of the scalar potential; that is, we may write
 \begin{eqnarray}
 \bmB_P & = & {\cal B} \cdot \bphi,
 \label{eqn:BPnmatrix}
 \end{eqnarray}
 where $\bmB_P$ represents the vector of Fourier harmonics of $B_{P,n}$, and ${\cal B}$ is a matrix derived from inserting the mixed Chebyshev-Fourier representation for $\Phi$ into \Eqn{BPn}.
 We note that ${\cal B}$ depends only on the geometry of $\pcV$ and $\cD$; i.e., ${\cal B}={\cal B}[\pcV,\cD]$.
 
 Provided that the external field $\bb_E$ is ``tame" (in particular $\displaystyle \int_{\cD} B_{E,n}\diff \bar s=0$), we put forward the following propositions: 
 (i) for toroidal annular domains of practical interest, with the constraint that $\bmB_T \cdot \bm{n} = 0$ on the inner boundary, $\pcV$, there exists a normal magnetic field on the outer boundary, $\cD$, that is the sum of an {\em a priori known} externally applied field plus the {\em a priori} {\em unknown} magnetic field that is produced by plasma currents inside $\pcV$, as given by the virtual-casing integral given; and (ii) that this ``virtual-casing self-consistent'' vacuum field may be obtained as the solution to the system of linear equations resulting from combining \Eqn{BPnmatrix} with \Eqn{Laplacematrix}; i.e.
 \begin{eqnarray}
{\cal L}_{vc}\cdot \bphi = \bmB_E,
 \label{eqn:selfconsistentvacuumfield}
 \end{eqnarray}
 where we defined the {\it Laplace-virtual-casing} matrix, ${\cal L}_{vc}\equiv{\cal A}-{\cal B}$, and $\bmB_E$ is that part of the right-hand-side vector given in \Eqn{Laplacematrix} that does not depend on the plasma currents.\footnote{In a private communication with SRH, Zhisong Qu independently suggested this approach.}
 We shall elaborate upon these propositions in a future article.
 
 The virtual-casing self-consistent vacuum field will change with variations in the plasma boundary according to
 \begin{eqnarray}
{\cal L}_{vc} \cdot \delta \bphi & = & - \left( \delta \cA - \delta {\cal B}\right)\cdot \bphi + \delta \bmB_E,
 \label{eqn:scvacuumfieldderivative}
 \end{eqnarray}
 where $\delta {\cal B} = \partial {\cal B} / \partial \pcV \cdot \delta \pcV + \partial {\cal B} / \partial \MB \cdot \delta \MB$.
 
 \subsection{Restricted energy functionals}
 \label{sec:reducedenergyfunctionals}

 To incorporate the Galerkin method for computing the magnetic fields with the energy functional and with the various combined plasma-coil optimization algorithms discussed in \Sec{optimization} below, we simplify as follows.
 We re-define the energy functionals, $\calF_v$, in the plasma volumes to
 \begin{eqnarray}
 \calF_v & \equiv & 
 \int_{{\cal{V}}_v} \left( \frac{p_v}{\gamma-1} + \frac{B_v^2}{2} \right) \diff v, 
 \label{eqn:Fvreduced}
 \end{eqnarray}
 where here and hereafter the magnetic field, $\bmB_v$, in each region is restricted to be the unique magnetic field with the prescribed helicity and fluxes that is tangential to the boundary and obeys $\nabla \times \bb_v = \mu_v \bb_v$. 
 This equation is known as the Beltrami equation and is obtained, see \Eqn{Beltrami}, as the Euler Lagrange equation $\delta \calF_V / \delta \bm{A}_v = 0$.
 We note that $\bmB_v$ depends only on the geometry of the adjacent interfaces; i.e., $\bmB_v = \bmB_v[\bmx_{v-1},\bmx_v]$.\footnote{The energy functional also depends on the mass and entropy profiles, the flux profiles, and the helicity profiles; but, in this article we assume that these are given. In practice, these must be chosen to be consistent with an assumed model of transport; or, so that the parallel current or rotational transform profiles match experimentally measured profiles, for example, or so that pressure gradients do not coincide with rational surfaces.}
 The energy functional in the vacuum region is redefined simply as
 \begin{eqnarray}
 \calF_+ & = & 
 \int_{\cV_+} \frac{B_+^2}{2}\, \diff v,
 \label{eqn:restrictedF+}
 \end{eqnarray}
 where $\bmB_+ = \nabla\Phi$ is restricted to be the unique magnetic field with the prescribed loop integrals (for the enclosed currents) and is tangential to $\pcV$, and that obeys $\nabla\cdot\nabla\Phi=0$ in $\calV_+$.
 This field is obtained as the solution to either \Eqn{Laplacematrix} if the total normal field is given or \Eqn{selfconsistentvacuumfield} if the external normal field is given.
 With the enclosed currents and $\cD$ held constant, we note that $\bmB_+$ depends only on ${\cal S}$ and the boundary condition; i.e., $\bmB_+ = \bmB_+[{\cal S},B_{T,n}]$ or $\bmB_+ = \bmB_+[{\cal S},B_{E,n}]$.

 With these conventions hereafter implied, we omit the explicit dependence of $\calF$ on the vector and scalar potentials and the Lagrange multipliers, and write $\calF=\calF_t[\bmx, B_{T,n},\MB]$ or $\calF=\calF_e[\bmx, B_{E,n},\MB]$. To distinguish the case when $\BTn$ is provided and \Eqn{Laplacematrix} is used to compute the vacuum field, we use $\calF_t$ to denote the energy functional in this case, whereas we instead denote this quantity by $\calF_e$ when $\BEn$ is provided and \Eqn{selfconsistentvacuumfield} is used.
 We use $\bmx$ to denote the geometry of the $v=1,\dots N_V$ interfaces, which includes the plasma boundary; i.e., $\bmx_{N_V} = \pcV$.
 
 The interface geometry, and by implication the equilibrium magnetic field, is defined by $\delta \calF / \delta \bmx = 0$, which from \Eqn{forcebalance} is the equilibrium equation, $\bm{F} \equiv [[p+B^2/2]]=0$ across the ideal interfaces.
 Minima of $\calF$ with respect to $\bm{x}$ can be found using descent-style algorithms, $\partial \bmx / \partial \tau = - \delta \calF/\delta \bmx$, where $\tau$ is an integration parameter, or by Newton-style methods.
 The geometry of the equilibrium interfaces is considered to be a function of the boundary conditions; i.e., $\bmx = \bmx[B_{T,n},\cD]$ or $\bmx = \bmx[B_{E,n},\cD]$.
 We shall write the total (tangential) magnetic field, $\bmB_T|_{\pcV}$, on $\pcV$, which is required for the virtual casing calculation of $B_{P,n}$, as a function of $\bmx$; i.e., $\bmB_T|_{\pcV}=\bmB_T|_{\pcV}[\bmx]$.
 
 The second derivatives of the restricted energy functions, \Eqn{Fvreduced} and \Eqn{restrictedF+}, with respect to variations in the interface geometry, $\bmx$, and the boundary conditions, either $B_{T,n}$ of $B_{E,n}$, which are required in the following section, may be constructed from incorporating the derivatives of the magnetic fields as calculated using \Eqn{vacuumfieldderivative} and \Eqn{scvacuumfieldderivative} with the expressions presented in \Sec{energyfunctionalvariations}.\footnote{All the required derivatives have been implemented in SPEC \citep{Hudson2012,Hudson2020}.}
 
 \section{Combined plasma-coil optimization algorithms}
 \label{sec:optimization}
 
 In this section, we consider the construction of the plasma equilibrium and the coil geometry simultaneously with the optimization of the plasma and coil properties.
 
 When we are required to construct the geometry of the coils as extrema of the coil-penalty functional, $\calE$, we assume that the coil geometry satisfies $\delta \calE / \delta \bmc = 0$.
 When the coil geometry is taken as the independent degree-of-freedom in the optimization, the coil-penalty functional is not required and the Biot-Savart integral is used to compute $\BEn = \BEn[\bmc,\cD]$.
 
 When the normal plasma field, $B_{P,n}$, on $\cD$ is required and the equilibrium magnetic field is known, $B_{P,n}$ is computed 
 using either the total (tangential) magnetic field either immediately inside, $\pcV_-$, or immediately outside, $\pcV_+$, the plasma boundary, depending on whether $\bmB_T$ is determined using a fixed-boundary calculation, for which the total magnetic field outside $\pcV$ may not be known, or a free-boundary calculation, for which it is.
 If there is a sheet current on the plasma boundary, these will differ.
 The total (tangential) magnetic field, $\bmB_T|_{\pcV}$ on $\pcV$, is obtained as an output of the equilibrium calculation, $\delta \calF / \delta \bmx = 0$.
 
 In the following, we consider choosing the independent degrees-of-freedom, $\bmz$, in the combined plasma-coil optimization to be 
 (i) the plasma boundary, $\bmz \equiv \pcV$;
 (ii) the total normal field on $\cD$, $\bmz \equiv B_{T,n}$; 
 (iii) the required normal field on $\cD$, $\bmz \equiv D_n$; and 
 (iv) the coil geometry, $\bmz \equiv \bmc$.
 For each of these choices, the descent direction depends on the derivatives of the plasma-energy functional, $\calF = \calF_t[\bmx_v, B_{T,n},\MB]$ or $\calF = \calF_e[\bmx_v, B_{E,n},\MB]$ depending on whether the total or external normal field is given; 
 the coil-penalty functional, $\calE = \calE[\bmc,D_n,\cD]$; 
 and the virtual-casing calculation of the plasma normal field, $B_{P,n} = B_{P,n}[\pcV,\bmB_T|_{\pcV},\cD]$.
 
 It is helpful to have a tangible example of what plasma and coil properties we wish to optimize.
 For the plasma optimization, we imagine a plasma property, ${\cal P}[\bmx]$, that is explicitly a scalar function of the geometry of the flux surfaces and is to be minimized.
 Generally, most plasma properties depend on the magnetic field, but here we are assuming that the equilibrium magnetic field depends on the geometry of the interfaces and so there is no loss of generality in treating ${\cal P}$ as an explicit function of only the geometry of the interfaces.
 Plasma properties of interest include the integrability of the magnetic field, quasi-symmetry properties, the rotational-transform profile, stability properties etc. 
 
 For the coil geometry optimization, we imagine a measure of the coil complexity, ${\cal C}[\bmc]$, that is a scalar function of the coil geometry and is to be minimized.
 We might consider the total length of the coils, as shorter coils tend to be cheaper, or the non-planar-ness of the coils, as measured by ${\cal C} = \sum_i \displaystyle \oint \tau^2/2 \diff l$
 where $\tau$ is the torsion.
 Other properties of interest might include a measure of the coil-plasma separation, which explicitly depends on both the coil geometry and the plasma boundary; e.g., ${\cal Q} = {\cal Q}[\bmx,\bmc]$. It is straightforward to extend the following treatment to include such properties.

 For the combined plasma-coil optimization, we imagine a differentiable {\it cost-function}, $\cal{T}=\cal{T}[\cal{P},\cal{C}]$. 
 If ${\cal P}[\bmx]$ and ${\cal C}[\bmc]$ are differentiable functions of, respectively, $\bmx$ and $\bmc$, then descent algorithms can be implemented if we know the derivatives of $\bmx$ and $\bmc$ with respect to $\bmz$. 
 The derivative of ${\cal T}$ is
 \begin{eqnarray}
 \frac{\partial \cal{T}}{\partial \bmz} 
 = 
 \frac{\partial \cal{T}}{\partial {\cal P}} 
 \frac{\partial {\cal P}}{\partial \bx}
 \frac{\partial \bx}{\partial \bmz} 
 + 
 \frac{\partial \cal{T}}{\partial {\cal C}} 
 \frac{\partial \cal{C}}{\partial \bmc}
 \frac{\partial\bmc}{\partial \bmz}. 
 \label{eqn:costfunction}
 \end{eqnarray}
 Here and hereafter, for notational clarity, we assume that all quantities are discretized, so that {\em functionals} of lines, surfaces and volumes become {\em functions} of a finite set of parameters that describe those objects, and the infinite-dimensional Frech\'et derivatives become finite-dimensional derivatives.
 One cannot be more explicit regarding constructing the derivatives of ${\cal T}$, ${\cal P}$ and ${\cal C}$ until one has stated what these quantities are, this is left to a future article.
 In the following, we present expressions for $\partial \bmx / \partial \bmz$ and $\partial \bmc / \partial \bmz$ for the different choices of $\bmz$ mentioned above.
 
 \subsection{Fixed-boundary optimization}
 \label{sec:fixedboundaryoptimization}
  
 We can consider the independent degree-of-freedom in the optimization to be the plasma boundary, $\bmz \equiv \pcV$.
 The free-plasma-boundary equilibrium calculation described in \Sec{threefunctionals} reduces to a fixed-plasma-boundary calculation by eliminating the vacuum region; that is, we choose $\MB=\pcV$ and $B_{T,n}=0$.
 In this subsection only, because we are choosing $\cD$ to be coincident with $\pcV$ to facilitate a fixed-boundary equilibrium calculation, $\cD$ will move during the optimization.
 
 The equilibrium, $\bmx$, satisfies $\partial \calF_t(\bmx,0,\pcV) / \partial \bmx=0$.
 We compute $B_{P,n}(\pcV,\bmB_T|_{\pcV_-},\pcV)$ from the virtual-casing integral.\footnote{If $\cD = \pcV$, there is a singularity in the integrand of the virtual casing equation. This can be overcome by subtracting an integral which posses the same type of singularity and adding the analytic expression of that subtracted integral \citep{MERKEL1986}.}
 With $B_{T,n}=0$, the required normal field is $D_n = - B_{P,n}(\pcV,\bmB_T|_{\pcV_-},\pcV)$.
 The coil geometry satisfies $\partial \calE(\bmc, D_n,\pcV)/\partial \bmc=0$. 
 Expanding and collecting terms, we obtain
 \begin{eqnarray} 
 \frac{\partial \bmx}{\partial \bmz} & = & - \left( \frac{\partial^2 \calF_t}{\partial \bmx \partial \bmx} \right)^{-1} \cdot \left( \frac{\partial^2 \calF_t}{\partial \bmx \partial \calD} \right), \\
 \frac{\partial \bmc}{\partial \bmz} & = & - \left( \frac{\partial^2 \calE}{\partial \bmc \partial \bmc} \right)^{-1} \cdot \left( \frac{\partial^2 \calE}{\partial \bmc \,\partial D_n} \cdot \frac{\partial D_n}{\partial \bmz} + \frac{\partial^2 \calE}{\partial \bmc \,\partial \calD} \right),
 \end{eqnarray}
 where
 \begin{eqnarray}
 \frac{\partial D_n}{\partial \bmz} & = & -  \frac{\partial \BPn}{\partial \bmx} \cdot \frac{\partial \bmx}{\partial \bmz},
 \end{eqnarray}
 and where 
 \begin{eqnarray}
 \frac{\partial \BPn}{\partial \bmx} & = & 
 \frac{\partial \BPn}{\partial \pcV}
 \cdot \frac{\partial \pcV}{\partial \bmx} 
 + 
 \frac{\partial \BPn}{\partial \bmB_T|_{\pcV}} \cdot \frac{\partial \bmB_T|_{\pcV_-}}{\partial \bmx}.
 \label{eqn:dBpndx-}
 \end{eqnarray}
 The analogous functional derivatives for $\partial^2 \calE / \partial \bmc \partial \bmc$ and $\partial^2 \calE / \partial \bmc \partial \cD$ are given in \Eqn{ddEdcdc} 
 and \Eqn{ddEdcdD}, those for $\partial^2 \calF_t/\partial \bmx \partial \bmx$, $\partial^2 \calF_t/\partial \bmx \partial \calD$ are given in \Eqn{ddFdAdxi} and \Eqn{secondvariationA}, and those for $\partial B_{P,n} / \partial \pcV$ and $\partial B_{P,n} / \partial \bmB_T|_{\pcV}$ can be found in  \Eqn{dBPndS}, and \Eqn{dBPndBT}.
 
 The coil geometry is constructed by minimizing the coil-penalty functional.
 With a finite number of discrete coils, the optimal coils will not exactly produce the required magnetic field.
 It may be beneficial for a combined plasma-coil optimization to minimize the quadratic-flux error $\varphi_2=\varphi_2(D_n,\bmc,\cD)$
 The required derivative is
 \begin{eqnarray}
 \frac{\partial \varphi_2}{\partial \bmz} & = & \frac{\partial \varphi_2}{\partial D_n} \cdot \frac{\partial D_n}{\partial \bmz} 
 +  
 \frac{\partial \varphi_2}{\partial \bmc} \cdot \frac{\partial \bmc}{\partial \bmz} + \frac{\partial \varphi_2}{\partial \cD}.
 \label{eqn:qferror}
 \end{eqnarray}
 The analogous functional derivative, $\partial \varphi_2/\partial \cD$, is given in \Eqn{dEdD}. 
 
 \subsection{Generalized fixed-boundary optimization}
 \label{sec:generalizedfixedboundaryoptimization}
 
 We can consider the independent degree-of-freedom in the optimization to be the total normal field on the computational boundary, ${\bmz} \equiv B_{T,n}$ on $\MB$.
 $\cD$ is assumed to lie outside the plasma boundary and to remain fixed during the calculation.
 The equilibrium satisfies $\partial\calF_t(\bmx,B_{T,n},\cD)/\partial\bmx=0$.
 The required normal field is $D_n = B_{T,n}-B_{P,n}(\pcV,\bmB_T|_{\pcV_+},\cD)$.
 The coil geometry satisfies $\partial \calE(\bmc,D_n,\cD)/\partial\bmc=0$. Expanding and collecting terms, we obtain
 \begin{eqnarray}
 \frac{\partial \bmx}{\partial \bmz} & = & - \left( \frac{\partial^2 \calF_t}{\partial \bmx \partial \bmx} \right)^{-1} \cdot \frac{\partial^2 \calF_t}{\partial \bmx \partial B_{T,n}}, \label{eqn:Adxdz} \\
 \frac{\partial \bmc}{\partial \bmz}  & = & - \left( \frac{\partial^2 \calE}{\partial \bmc \partial \bmc} \right)^{-1} \cdot \frac{\partial \calE}{\partial \bmc \partial D_n} \cdot \frac{\partial D_n}{\partial \bmz}, \label{eqn:Adcdz}
 \end{eqnarray}
 where
 \begin{eqnarray}
 \frac{\partial D_n}{\partial \bmz} & = & 1 - \frac{\partial \BPn}{\partial \bmx} \cdot \frac{\partial \bmx}{\partial \bmz},
 \end{eqnarray}
 and 
 where 
 \begin{eqnarray}
 \frac{\partial \BPn}{\partial \bmx} & = & 
 \frac{\partial \BPn}{\partial \pcV} \cdot \frac{\partial \pcV}{\partial \bmx} 
 + 
 \frac{\partial \BPn}{\partial \bmB_T|_{\pcV}} \cdot \frac{\partial \bmB_T|_{\pcV_+}}{\partial \bmx}.
 \label{eqn:dBpndx+}
 \end{eqnarray}
  The analogous functional derivatives for $\partial \calE/\partial \bmc \partial D_n$, $\partial B_{P,n} / \partial \pcV$, and $\partial B_{P,n} / \partial \bmB_T|_{\pcV}$ are given in \Eqn{ddEdcdDn}, \Eqn{dBPndS}, and \Eqn{dBPndBT}.
 Note the use of $\bmB_T|_{\pcV_-}$ in \Eqn{dBpndx-} and $\bmB_T|_{\pcV_+}$ in \Eqn{dBpndx+}.
 The derivative of the quadratic-flux error is
 \begin{eqnarray}
 \frac{\partial \varphi_2}{\partial \bmz} =   \frac{\partial \varphi_2}{\partial D_n} \cdot \left( 1 - \frac{\partial B_{p,n}}{\partial \bmx} \cdot \frac{\partial \bmx}{\partial \bmz} \right)
  + \frac{\partial \varphi_2}{\partial \bmc} \cdot \frac{\partial \bmc}{\partial \bmz}.
 \end{eqnarray}
 The analogous functional derivative for $\partial \phi_2 /\partial D_n$ is given in \Eqn{dEdDn}.
 
 \subsection{Quasi-free-boundary optimization}
 \label{sec:quasifreeboundaryoptimization}
 
 In this case, we consider the independent degree-of-freedom in the optimization to be the required external normal field on the computational boundary, $\bmz \equiv D_n$ on $\cD$.
 For the equilibrium calculation, we assume that the ``actual'' external normal field is equal to the required external normal field, $\BEn = D_n$; i.e., we assume that the equilibrium satisfies $\partial\calF_e(\bmx,D_n,\cD)/\partial\bmx=0$.
 The virtual casing integral is not explicitly required as it is implicitly included in the construction of the virtual-casing self-consistent vacuum field, \Eqn{selfconsistentvacuumfield}.
 The coil geometry satisfies $\partial \calE(\bmc,D_n,\cD) / \partial\bmc = 0$.
 
 Expanding and collecting terms, we obtain
 \begin{eqnarray}
 \frac{\partial \bmx}{\partial \bmz} & = & - \left( \frac{\partial^2 \calF_e}{\partial \bmx \partial \bmx} \right)^{-1} \cdot \left( \frac{\partial^2 \calF_e}{\partial \bmx \partial B_{E,n}} \right), \\
 \frac{\partial \bmc}{\partial \bmz} & = & - \left( \frac{\partial^2 \calE}{\partial \bmc \partial \bmc} \right)^{-1} \cdot \left(  \frac{\partial^2 \calE}{\partial \bmc \partial D_n} \right),
 \end{eqnarray}
 and for the quadratic-flux error
 \begin{eqnarray}
 \frac{\partial \varphi_2}{\partial \bmz} & = &
 \frac{\partial \varphi_2}{\partial D_n} + \frac{\partial \varphi_2}{\partial \bmc} \cdot \frac{\partial \bmc}{\partial \bmz}.
 \end{eqnarray}

 \subsection{Free-boundary (coil) optimization} 
 \label{sec:freeboundaryoptimization}
  
 In this final case, the independent degree-of-freedom in the optimization is the coil geometry, $\bmz \equiv \bmc$.
 The equilibrium satisfies $\partial \calF_e(\bmx,B_{E,n},\cD)/\partial \bmx=0$, where $B_{E,n}=B_{E,n}[\bmc,\cD]$ is computed using the Biot-Savart integral.
 
 We obtain
 \begin{eqnarray}
 \frac{\partial \bmx}{\partial \bmz} & = & - \left( \frac{\partial^2 \calF_e}{\partial \bmx \partial \bmx} \right)^{-1} \cdot \frac{\partial^2 \calF_e}{\partial \bmx \partial B_{E,n}} \cdot \frac{\partial B_{E,n}}{\partial \bmc}.
 \end{eqnarray}
 The analogous functional derivative for $\partial B_{E,n}/\partial \bmc$ is given in \Eqn{dBndc}. 
 The coil geometry is given directly, $\bmc = \bmz$.
 
 Since the input to the equilibrium calculation, $\BEn$, is computed directly from the actual coils there is no coil error, $\varphi_2 = 0$.
 
 \section{Discussion}
 \label{sec:discussion}

 In this paper, we have categorized and investigated four different combined plasma-coil optimization approaches and proposed a novel method for improving free-boundary equilibrium calculation. 
 We began by summarizing all the functional derivatives of the MRxMHD energy,%
 \footnote{We could have used the ideal MHD functional instead of the MRxMHD functional. In fact, we believe that this would have simplified some of the algebra; however, we chose MRxMHD because it can treat rational surfaces better than ideal MHD.} the coil-penalty, and the virtual-casing integral needed for a combined plasma-coil optimization. 
 We emphasized that absence of an energy-principle formulation with an explicit boundary condition on the normal component of the external magnetic field on the computational boundary if it does not coincide with the plasma boundary. 
 This construction of the field in the vacuum region is advantageous for the free-boundary optimization approach. 
 For this special case, we proposed solving for the required field using a weak formulation. 
 Finally, we explicitly stated which derivatives are necessary for the four distinct combined plasma-coil optimization algorithms: fixed-boundary optimization, generalized fixed-boundary optimization, quasi-free-boundary optimization, and free-boundary (coil) optimization. To the best of our knowledge, all existing stellarator optimization algorithms can be grouped in either the fixed-boundary optimization or the free-boundary optimization approach.
 The collection of these four distinct optimization algorithms helps to clarify certain intrinsic problems of combined plasma-coil optimization, as we will discuss later in this section.
 
 The novel proposed approach for calculating the virtual-casing self-consistent vacuum field will reduce the cost of the free-boundary calculation to something comparable to that of a fixed-boundary calculation.  
 A full-Newton method, with analytic derivatives, can be used for both. 
 In the proposed approach, in \Eqn{selfconsistentvacuumfield} appears the Laplace-virtual-casing matrix ${\cal L}_{vc}={\cal A}-{\cal B}$, which is an important matrix that deserves some discussion. 
 This matrix, in a vague sense, ``connects'' the plasma to the external magnetic field. 
 Since it is not symmetric, it cannot directly follow from an energy principle, although there are round-about ways to force this to happen (e.g. by introducing adjoint degrees of freedom). Investigation of the well-posedness of the Laplace-virtual-casing problem is the subject of future publication. 

 On the basis of the summary of the four distinct optimization algorithms presented in \Sec{optimization}, we now discuss the differences and potential advantages and disadvantages of the various optimization algorithms described above. 
The independent degree-of-freedom is 
 (i) a two-dimensional surface, $\pcV(\theta,\zeta)$, where $\theta$ and $\zeta$ are poloidal and toroidal angles, embedded in three-dimensional space for the fixed-boundary approach; 
 (ii) a scalar function, $\BTn(\theta,\zeta)$, with the constraint that the net-flux is zero for the generalized fixed-boundary approach; 
 (iii) a similar function, $D_n(\theta,\zeta)$, for the quasi free-boundary approach; and 
 (iv) a discrete set of one-dimensional curves embedded in three-dimensional space, $\bmc_i(\theta)$ for $i = 1, \dots, N_C$, for the free-boundary (coil) approach.
 Note that a surface is really just one scalar function of two angles: even though some codes represent a surface using two functions, namely $R$ and $Z$, this introduces a tangential null space, which is removed by exploiting spectral condensation \citep{SpectralCondenation1,Lee1988,SpectralCondensation3}. 
 Also, in (iv) we have restricted attention to when the external field is provided by a finite number of closed current filaments; other representations can be included.
 
 For the quasi free-boundary approach, the theory of ``efficient fields'' can be used when $D_{n}$ is the independent degree-of-freedom ~\citep{Landreman2016} to determine which Fourier spectrum of the normal component of the magnetic field corresponds to distant coils. 
 Using only efficient fields, the optimization space can be effectively reduced.
 For the free-boundary (coil) algorithm, we expect that a concise parameterization of the coils can be introduced.
 For the other approaches, we suggest to use some measure of the coil error in total cost function: the quadratic-flux error would be a good target, and targets weighting resonant normal field errors are probably better.

 In each presented case, the $\partial^2{\cal F}/\partial \bmx\partial \bmx$ matrix needs to be inverted to calculate $\partial \bmx / \partial \bmz$. 
 This is the equilibrium stability matrix, and near marginal stability its eigenvalues vanish. 
 This suggests that the optimization can encounter singular points. 
 Physically, this has to do with the fact that if a plasma is only marginally stable the equilibrium can be change substantially by even a small perturbation. 
 This presumably would induce a large change in the coil geometry, whose optimal shape is therefore very uncertain. 
 Marginal stability is a particularly relevant and important case. 
 Toroidal confinement of fusion plasmas typically improves as the plasma pressure and or currents increase, but pressure and currents ultimately are responsible for instabilities. 
 An economically viable fusion reactor would presumably operate at the highest fusion-performance parameters but also sufficiently far from bifurcation boundaries in parameter space so that small variations could be controlled. 
 
 Similarly, in all but the last case, the $\partial^2 \calE/\partial\bmc\partial\bmc$ matrix needs to be inverted to calculate $\partial\bmc/\partial \bmz$. 
 Small changes in the equilibrium may result in large changes to the coils if this matrix has a small eigenvalue.
 
 Algorithms for combined plasma-coil optimization might become problematic near marginal stability boundaries, particularly if the response in either the plasma or the coils to small variations is not well understood.
 Depending on the plasma and coil properties chosen, there might be bifurcation boundaries in the optimization parameter space, which determine which initial conditions will converge to which local minima.

 Marginal plasma stability is a lower-dimensional subspace of the full space of configurations, and it can be defined when there is a direction in which there is no variation, i.e. there are eigenvalues that are zeros.
 A similarly defined marginal stability subspace exists in the coil space.
 The stability boundary of the combined plasma-coil optimization, in the parameter space denoted by $\bm{z}$, will be a combination of both.
 Depending on the plasma, ${\cal P}(\bm{x})$, and coil, ${\cal C}(\bm{c})$, properties chosen, there may exist subspaces for which $\partial {\cal P} / \partial \bm{x}=0$ and $\partial {\cal C} / \partial \bm{c}=0$, and there may be additional bifurcation boundaries in ${\cal T}(\bm{z})$, depending on what ${\cal T}(\bm{z})$ is chosen to be. 
 For robust multi-objective functional optimization, it would be advantageous to understand separately the stability boundaries in the plasma, coil and optimization spaces for a variety of relevant plasma and coil properties.
 
 In this paper, we focused on the $\partial \bmx/\partial \bmz$  and $\partial \bmc/\partial \bmz$  part of \Eqn{costfunction}, because these derivatives appear independently of which properties one wants to tailor in the optimization.  
 We neglected the actual cost-function with the equilibrium, ${\cal P}$, and coil, ${\cal C}$, properties and their derivatives because these quantities are problem specific; for example, a reactor-grade plasma has different design criteria than a research experiment.
 Recently, much work has been devoted to the development of adjoint methods and automatic differentiation to calculate derivatives of the plasma and coil properties, e.g. for the rotational transform, neoclassical transport and for the volume averaged $\beta$ \citep{Adjoint1,Adjoint2,Adjoint3,Adjoint4,Adjoint5}.
 These methods promise significant advantages for stellarator optimization.

 For both the equilibrium calculation and the coil calculation, both descent- and Newton-style methods can be used.
 Descent methods can be employed for combined plasma-coil optimization.
 An interesting question is whether or not all these descent calculations can be performed simultaneously; for example,
 \begin{eqnarray}
 \frac{\partial \bmx}{\partial \tau} = - \alpha \frac{\partial \calF}{\partial \bmx}, \;\;\;\;
 \frac{\partial \bmc}{\partial \tau} = - \beta  \frac{\partial \calE}{\partial \bmc}, \;\;\;\;
 \frac{\partial \bmz}{\partial \tau} = - \frac{\partial {\cal T}}{\partial \bmz}, \label{eqn:integrated}
 \end{eqnarray}
 where $\tau$ is an arbitrary integration parameter, and $\alpha$ and $\beta$ can be chosen for numerical stability so that, for example, the ``inner'' equilibrium and coil geometry calculations are sufficiently converged so that the ``outer'' optimization calculation is provided with sufficiently accurate information. 
 The equations in \Eqn{integrated} can be thought of as a dynamical system; and, like most dynamical systems, the location and stability of the fixed points of \Eqn{integrated} gives important information about the ``dynamics'' \citep{DynamicalSystems}, which in this case is the convergence and stability properties of the combined plasma-coil optimization.
 
 \section{Acknowledgements}
 The lead author would like to thank J. Loizu, E. Paul, and B. Shanahan for helpful conversations.
 This work has been carried out in the framework of the EUROfusion Consortium and has received funding from the Euratom research and training programme 2014-2018 and 2019-2020 under grant agreement no. 633053. It was also supported by a grant from the Simons Foundation (560651, PH).
 The views and opinions expressed herein do not necessarily reflect those of the European Commission.

 \bibliographystyle{jpp}

 \bibliography{Ref}



 \end{document}